\documentclass[12pt,english,preprint]{elsarticle}
\usepackage[T1]{fontenc}
\usepackage[utf8]{inputenc}
\usepackage{color}
\usepackage{babel}
\usepackage{amstext}
\usepackage[bookmarks=false,
 breaklinks=true,pdfborder={0 0 0},pdfborderstyle={},backref=false,colorlinks=true]
 {hyperref}

\makeatletter

\newcommand{\noun}[1]{\textsc{#1}}

\newcommand{\pt}{\ensuremath{p_{\rm{T}}}}

\makeatother

\begin{document}
\begin{frontmatter}
\title{Physics Objects in CMS Run 3\tnoteref{copyright}}
\tnotetext[copyright]{Copyright 2026 CERN for the benefit of the CMS Collaboration. CC-BY-4.0
license.}
\author{Markus Seidel on behalf of the CMS Collaboration}
\ead{markus.seidel@cern.ch}
\address{Riga Technical University, Riga, Latvia}
\begin{abstract}
In these proceedings we review the physics objects used by the CMS
experiment during LHC Run~3 at $\sqrt{s}=13.6$ TeV, including charged
leptons, photons, jets, and missing transverse momentum. Their performance
and calibration is critical for physics analysis. In particular, the
algorithms need to be resilient against the high pileup conditions
in Run~3 collisions. Furthermore, transformer-based algorithms are
deployed for the identification of heavy-flavor jets and boosted resonances.
\end{abstract}
\end{frontmatter}

\section{Introduction}

The CMS experiment~\cite{CMS:2008xjf} is currently operating in
Run~3 of the Large Hadron Collider (LHC). The accelerator has delivered
over 500~fb$^{-1}$ of data across Runs~1, 2, and 3 combined, with
the CMS detector successfully recording these collisions. A key characteristic
of the current data-taking period is the high instantaneous luminosity,
which results in around 60 proton-proton interactions per bunch crossing
(pileup) on average. To maintain sensitivity to rare processes and
precision measurements, the CMS Collaboration has developed pileup-resistant
reconstruction algorithms.

The precise determination of the luminosity~\cite{CMS:2021xjt} is
fundamental for absolute cross-section measurements. For the 2023
Run~3 data, the preliminary uncertainty is estimated at $\pm1.3\%$
and expected to be further improved. Furthermore, a refined analysis
of the full Run~2 dataset (2016--2018) has resulted in a total integrated
luminosity of $137.88\pm1.01$ fb$^{-1}$, corresponding to an uncertainty
of $0.73\%$~\cite{CMS-PAS-LUM-20-001}. This represents a significant
improvement over the 1.75\% uncertainty used in previous top quark
cross-section measurements.

The CMS event reconstruction relies on the Particle Flow (PF) algorithm~\cite{CMS:2017yfk},
which identifies muons, electrons, photons, and charged and neutral
hadrons by combining information from the tracker, calorimeters, and
the muon system. Double-counting of energy is prevented by subtracting
track momenta from calorimeter clusters.

A novel development for Run 3 is the Machine Learning Particle Flow
(MLPF) algorithm~\cite{CMS-DP-2025-033}. MLPF utilizes a transformer
model to infer the full particle content directly from tracks and
clusters. Trained on $t\bar{t}$, dijet, and $Z\to\tau_{h}\tau_{h}$
events, MLPF demonstrates comparable physics performance and achieves
a factor 2 speedup on modern heterogeneous computing infrastructure
(CPU+GPU) compared to PF.

\section{Lepton and Photon Reconstruction and Calibration}

\subsection{Muons}

Muons are reconstructed by linking tracks in the silicon tracker with
segments and hits in the muon system~\cite{CMS:2018rym}. For Run
3, the isolated single muon trigger threshold remains at 24~GeV.
This serves as a multi-purpose trigger, with isolation requirements
relaxed during 2024 to increase efficiency~\cite{CMS-DP-2025-014}.
Muon identification relies on both cut-based criteria and multivariate
analysis techniques~\cite{CMS:2023dsu}, and specialized algorithms
are used for both low~\cite{CMS:2024qyo} and high $\pt$~\cite{CMS:2019ied}
muons. Efficiencies are monitored throughout Run~3~\cite{CMS-DP-2024-067}
using the tag-and-probe method on $J/\Psi$ and $Z\to\mu\mu$ resonances~\cite{CMS:2024qgs}.

Momentum corrections are applied to address magnetic field mismodeling
(multiplicative corrections) and tracker alignment biases (additive
corrections). The standard calibration, which is available for the
2024 dataset, achieves a precision of approximately 0.05\%~\cite{CMS-DP-2024-065},
while dedicated analyses such as the W mass measurement employ custom
track fits to reach a precision of 0.006\%~\cite{CMS:2024lrd}.

\subsection{Electrons and Photons}

Electrons are reconstructed using the Gaussian Sum Filter (GSF) algorithm
to account for bremsstrahlung energy loss, combined with ``mustache''
superclusters in the ECAL. An energy regression based on shower shape
and pileup density improves the electron response and resolution~\cite{CMS:2020uim}.
Electron efficiencies and energy corrections were derived from $Z\to ee$
events~\cite{CMS-DP-2024-052}. A new tag-and-probe method based on
unbinned likelihood fits and normalizing flows has been explored for
Run~3~\cite{CMS-DP-2025-053}. This allows for the derivation of
high-dimensional efficiency maps and event-by-event scale factors.

Photon identification relies on shower shapes and isolation variables.
The MVA-based photon ID has been improved for Run 3 to ensure stability
against pileup. The energy scale was calibrated using $Z\to ee$ events
and validated in $Z\to\mu\mu\gamma$ events~\cite{CMS-DP-2024-052}.

\section{Jet Reconstruction and Calibration}

Jets are clustered from PF candidates using the anti-$k_{t}$ algorithm
with $R=0.4$ and $0.8$. The calibration chain comprises of the L1
pileup offset correction and the simulation-based response corrections,
followed by residual data/simulation corrections derived from dijet
(relative $\eta$ calibration) and $\gamma/Z$+jet events (absolute
$p_{T}$ calibration)~\cite{CMS:2016lmd}. For Run~2, the total
uncertainty of the jet energy calibration is at the level of 1\% (excluding
time and flavor-dependent effects), while the jet energy resolution
in MC requires a correction of $+10\pm4\%$~\cite{CMS-DP-2021-033}
to match the one in data.

For Run 3, the CMS Collaboration fully adopted and optimized the Pileup
Per Particle Identification (PUPPI) algorithm as the default for particle-based
pileup mitigation~\cite{Bertolini:2014bba,CMS-DP-2024-043}. Unlike
the Charged Hadron Subtraction (CHS) method used previously, PUPPI
uses local shape information to rescale particle four-momenta, minimizing
pileup contributions. PUPPI was shown to render boosted object identification
more resilient against pileup and strongly suppress pileup jets within
the tracker coverage~\cite{CMS:2020ebo}.

In addition, the pileup energy offset for PUPPI jets is close to zero,
rendering the traditional L1 pileup offset correction largely unnecessary~\cite{CMS-DP-2024-039}.
Calibrations have been derived for the 2022, 2023, and 2024 datasets~\cite{CMS-DP-2024-039,CMS-DP-2025-042}.
The preliminary 2025 corrections show improvements in the endcap due
to better calorimeter calibration~\cite{CMS-DP-2025-044}.

\section{Heavy Flavor and Boosted Object Tagging}

The identification of heavy-flavor jets has evolved from Recurrent
Neural Networks such as DeepJet~\cite{Bols:2020bkb} (used in Run~2)
to transformer-based architectures:
\begin{itemize}
\item ParticleNet: This graph neural network treats jet constituents as
an unordered ``particle cloud'', allowing it to better incorporate
low-level jet information~\cite{Qu:2019gqs}. 
\item UParT (Unified Particle Transformer): This model leverages pairwise
interaction features and includes adversarial training to ensure robustness
against simulation mismodeling. It enables strange-jet tagging for
the first time and includes quark/gluon discrimination~\cite{Qu:2022mxj,CMS-DP-2024-066,CMS-DP-2025-081}.
\end{itemize}
The new algorithms have been successfully commisioned in Run~3~~\cite{CMS-DP-2024-024}.
In addition, these advanced architectures support jet $p_{T}$ regression
as an auxiliary task, improving the jet energy resolution by up to
20\% after MC corrections~\cite{CMS-DP-2024-064}.

Tau lepton candidates are seeded using the hadrons-plus-strips algorithm~\cite{CMS:2018jrd}
and identified by the DeepTau algorithm~\cite{CMS:2022prd}. DeepTau~v2.5~\cite{CMS:2025kgf}
was used for the Run~2 reprocessing and Run~3 and has 30--50\%
lower jet mis-identification probability than the previous v2.1. In
addition, tau leptons can be reconstructed using the default jet algorithms
(PF+CHS or PF+PUPPI) and identified using ParticleNet or UParT. The
performance of all algorithms has been compared~\cite{CMS-DP-2025-073}
and validated in data~\cite{CMS-DP-2025-074}.

For boosted top quark identification, ParticleNet and ParT are applied
to large-radius AK8 jets~\cite{CMS:2025kje,CMS-PAS-JME-25-001}. These
taggers provide up to 10 times better background rejection than the
mass-decorrelated DeepAK8 tagger used in previous runs. They show
improved top quark mass resolution with regard to the soft drop algorithm~\cite{Larkoski:2014wba}
and utilize flat sampling of training samples to prevent mass sculpting
of the QCD background.

\section{Missing Transverse Momentum}

The reconstruction of missing transverse momentum ($p_{T}^{\text{miss}}$)
has also benefited from the transition to PUPPI~\cite{CMS:2019ctu}.
While PF $p_{T}^{\text{miss}}$ (without pileup subtraction) was the
default in Run~2, PUPPI $p_{T}^{\text{miss}}$ is recommended for
Run~3 due to its reduced dependence on pileup and improved resolution.
Additionally, the DeepMET algorithm based on deep neural networks
has been developed, offering a 10--30\% improvement in resolution
by learning the optimal weight for each individual PF candidate to
estimate the $p_{T}^{\text{miss}}$~\cite{CMS:2025prt}.

\section{Monte Carlo Modeling}

Precision measurements require accurate simulation of signal and background
processes. The standard $t\bar{t}$ simulation for Run~2 and 3 relies
on the \noun{Powheg-hvq} generator~\cite{Frixione:2007nw} interfaced
with \noun{Pythia~8}~\cite{Sjostrand:2014zea} using the CP5 tune~\cite{CMS:2019csb}.
The improvements in $t\bar{t}$ modeling were summarized in~\cite{CMS:2024irj}
and include:
\begin{itemize}
\item Initial and final state radiation: The modeling of jet multiplicity
and jet substructure in $t\bar{t}$ events is improved with respect
to Run~1, and event weights are used to estimate the uncertainties
with better statistical precision~\cite{Mrenna:2016sih}.
\item Top $p_{T}$: The Run~2 and 3 simulations show significantly better
agreement with data compared to the Run~1 simulation, with further
improvements expected from \noun{MiNNLO} generators~\cite{Mazzitelli:2021mmm}.
\item Color Reconnection (CR): Multiple CR models have been tuned to CMS
data~\cite{CMS:2022awf}, including the QCD-inspired model~\cite{Christiansen:2015yqa}
which better describe the ATLAS color flow data~\cite{ATLAS:2015ytt}.
Notably, the models only show differences when the top quark is allowed
to decay early for its decay products to take part in the CR process.
\end{itemize}

\section{Summary}

The physics objects used for CMS analysis (jets, leptons, and $p_{T}^{\text{miss}}$)
have successfully been commissioned and calibrated for the high-pileup
environment of LHC Run~3. The widespread adoption of the PUPPI algorithm
and the integration of transformer-based tagging represent significantly
improve performance despite harsher conditions. Combined with precise
luminosity determination and refined Monte Carlo modeling, CMS is
well-positioned to continue delivering precision top quark physics
results in the coming years. 

\bibliographystyle{elsarticle-num}
\bibliography{top2025_mseidel}

\end{document}